\documentclass[fleqn,twoside]{article}
\usepackage{espcrc2}

\usepackage{graphicx}
\usepackage[figuresright]{rotating}
\newcommand{\AmS}{{\protect\the\textfont2
  A\kern-.1667em\lower.5ex\hbox{M}\kern-.125emS}}

\title{ Angra Neutrino Project: status and plans\thanks{Talk given by J.C. Anjos
(janjos@cbpf.br) at NuFact05 Workshop, 21-26 June 2005, Frascati,
Italy. This Project is sponsored by LAMPADIA Foundation.}}

\author{J.C. Anjos\address[CBPF]{Centro Brasileiro de Pesquisas F\'{i}sicas,
        Rua Xavier Sigaud 150, Rio de Janeiro, RJ, 22290-180, Brazil},
A.F. Barbosa\addressmark[CBPF],
R.Z. Funchal\address[USP]{Instituto de F\'{i}sica, Universidade de
S\~{a}o Paulo, Caixa Postal 66318 S\~{a}o Paulo, SP, 05315-970, Brazil},
E. Kemp\address[UNICAMP]{Instituto de F\'{i}sica Gleb Wataghin, Universidade Estadual de Campinas,
Caixa Postal 6165, Campinas, SP, 13083-970, Brazil}, J.
Magnin\addressmark[CBPF], H. Nunokawa\address[PUC]{Departamento de
F\'{i}sica, Pontif\'{i}cia Universidade Cat\'{o}lica do Rio de
Janeiro, Caixa Postal 38071, Rio de Janeiro, RJ, 22452-970,
Brazil}, O.L.G. Peres\addressmark[UNICAMP],
D. Reyna\address[ARGONNE]{Argonne National Laboratory,
9700 S. Cass Avenue, Argonne, IL, 60439, USA}, R.C.
Shellard\addressmark[CBPF]}

\begin{document}

\begin{abstract}
We present the status and plans of the Angra Project, a new
nuclear reactor neutrino oscillation experiment, proposed to be
built in Brazil at the Angra dos Reis nuclear reactor complex.
This experiment is aimed to measure $\theta_{13}$, the last
unknown of the three neutrino mixing angles. Combining a high
luminosity design, very low background from cosmic rays and
careful control of systematic errors at the 1\% level, we propose
a high sensitivity multi-detector experiment, able to reach a
sensitivity to antineutrino disappearance down to $\sin^2
2\theta_{13} = 0.006$ in a three years running period, improving
present limits constrained by the CHOOZ experiment by more than an
order of magnitude.

\vspace{1pc}
\end{abstract}

\maketitle

\section{Introduction}

The neutrino oscillation phenomena still depends on three
unknown parameters: mixing angle $\theta_{13}$, sign of
$\Delta m^2_{13}$, and the CP phase $\delta$. Measurement of
$\theta_{13}$ in appearance experiments, such as the ones based
on accelerator, are subjected to suffer from the so called
parameter degeneracies due to the effect of the unknown sign
of $\Delta m^2_{13}$ and the CP phase $\delta$ in neutrino
oscillations~\cite{Hiroshi}.

%The neutrino oscillation matrix still has two unknown parameters:
%the mixing angle $\theta_{13}$ and the CP phase $\delta$.
%Appearance experiments, such ones based on accelerator beams, are
%subject to degeneracies in $\theta_{13}$ measurements due to the
%CP phase effect on the observed oscillations
The reactor neutrino short baseline experiments can produce a clear
oscillation signal and measure $\theta_{13}$ with no ambiguities
or matter effects~\cite{Whitepaper}. In this paper we describe the
main features and parameters of the Angra Experiment, a project to
measure antineutrino disappearance at the Angra dos Reis Nuclear
Complex (RJ - Brazil).

\section{Angra dos Reis: The reactor complex and site main features}

Angra dos Reis is a city located about 150 km south of Rio de
Janeiro. At 30 km from the city there is the nuclear complex,
having two operational reactors (Angra-I and II). The state owned
company Eletronuclear is responsible for the general management
and commercial operation of the plant. The thermal power of the
reactors are 2 GW and 4 GW for Angra I and II, with uptime around
83\% and 90\%, respectively.  The topology of the surrounding
terrain, formed by mountainous granite, is an advantage of the
site. The required overburden can be achieved by construction of
horizontal tunnels with lower cost than vertical shafts.

%Civil construction cost can be reduced by building horizontal
%tunnels in order to obtain substantial overburden for the
%detectors.

\section{The experimental design}

The Angra experiment will consist of 2 neutrino detectors in the
standard near/far configuration. The detectors should be built in
the 3 volume design: i) the central target filled with liquid
scintillator doped with gadolinium; ii) the surrounding volume
filled of standard scintillator (gamma catcher) and iii) the
non-scintillating buffer, with same optical properties of the
innermost zones, shielding the radioactive from outside. The
photo-tubes will be installed on the outer wall of the buffer. The
near site location is 300 m from the reactor core. We plan to
place a 50 ton target detector in a 100 m depth shaft providing
250 m.w.e. of overburden. At a distance of 1.5 km from the reactors
there is a 700 m granite peak of the so called "Morro do Frade".
The installation of the 500 ton far detector under this peak (2000
m.w.e.) provides an effective combination of detector distance and
overburden that increases the expected signal to noise ratio. In
addition, we intend to include a 1 ton very near detector, at 50 m
from the core, for flux precise monitoring, accurate spectral
shape measurement and general cross checks.

\section{Experimental Reach}

Preliminary estimations of the signal and
background rates and are shown in Table 1.
\vspace*{-0.6cm}
\begin{table}[htp]
%\begin{table*}[htp]
\newcommand{\m}{\hphantom{$-$}}
\newcommand{\cc}[1]{\multicolumn{1}{c}{#1}}
\renewcommand{\tabcolsep}{0.70pc} % enlarge column spacing
\renewcommand{\arraystretch}{0.9} % enlarge line spacing
\begin{tabular}{@{}llll}
\hline
Detector                & \cc{Very Near} & \cc{Near} & \cc{Far}  \\
\hline
Signal                  & \m1800 & \m2500 & \m1000 \\
Muons (Hz)              & \m150 & \m$\sim30$ & \m0.3  \\
$^{9}$Li bkg            & \m44  & \m $\leq20$  & \m $\sim2$ \\
\hline
\end{tabular}%\\[7pt]
%\end{table*}
\caption{The Angra Experiment expected rates. Signal and $^{9}$Li
background (correlated noise) are in events/day units.} \label{table:1}
\end{table}

The expected sensitivity as function of integrated luminosity is
shown in Figure 1. The assumed value of $\Delta m_{13}^{2}$ is
indicated in the figure. The calculations were performed by
minimizing the $\chi^{2}$ formulae built to take into account four
different types of systematic errors $\sigma_{ij}$ with assumed
values indicated in the figure. The subscripts D(d) represents
errors correlated (uncorrelated) between detectors and B(b) errors
correlated (uncorrelated) between bins of the measured energy
spectra \cite{Minakata}. As can be seen a limit of $\sin^2
2\theta_{13} = 0.006$ at 90\% confidence level can be achieved
within three years.

\section{Status and plans}

We are currently developing the very near detector, that will
serve as prototype to test detector elements and performance and
also as survey tool for systematic studies. This detector will
also be used to monitor the reactor activity, and to provide an
additional tool on verification of safeguards on
Non-Proliferation.
%\cite{Bowden}.
The planned turn-on dates are 2008 for the very
near detector and 2013 for the Angra complete configuration.
%\vspace*{-0.2cm}

\begin{figure}[htb]
\vglue -0.5cm
\includegraphics[height=6.6cm, width=7.5cm]{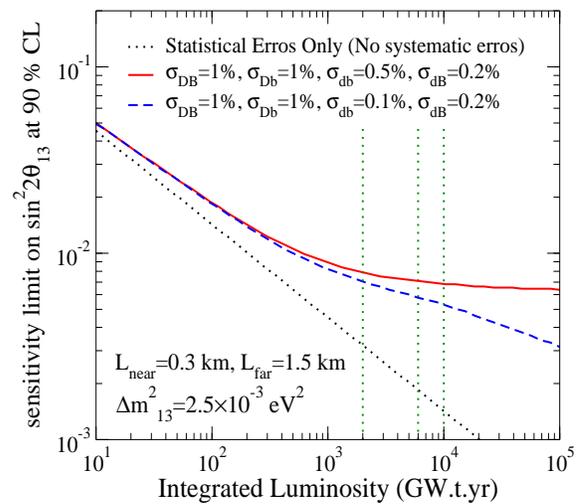}
\vglue -0.7cm
\caption{Sensitivity limit as a function of integrated luminosity.
Vertical dashed lines represent 1, 3 and 5 years of data taking.}
\label{fig:sensitivitylimit}
\end{figure}

\end{document}